\documentclass[]{spie}  

\usepackage{amsmath,amsfonts,amssymb}
\usepackage{graphicx}
\usepackage[font=bf]{caption}
\usepackage[colorlinks=true, allcolors=blue]{hyperref}
\usepackage[table]{xcolor}
\definecolor{light-gray}{gray}{0.95}

\title{A Multi-Scale Conditional Deep Model \\ for Tumor Cell Ratio Counting}

\author[a]{*Eric Cosatto}
\author[a]{Kyle Gerard}
\author[a]{Hans-Peter Graf}
\author[b]{Maki Ogura}
\author[b]{Tomoharu Kiyuna}
\author[c,d,e]{Kanako C Hatanaka}
\author[f,e]{Yoshihiro Matsuno}
\author[d,e]{Yutaka Hatanaka}

\affil[a]{Dept. of Machine Learning, NEC Laboratories America, Princeton, NJ, USA}
\affil[b]{Digital Healthcare Business Development Office, NEC Corporation, Tokyo, Japan}

\affil[c]{Clinical Research and Medical Innovation Center \protect\\ Hokkaido University Hospital, Hokkaido, Japan}
\affil[d]{Research Division of Genome Companion Diagnostics \protect\\ Hokkaido University Hospital, Hokkaido, Japan}
\affil[e]{Center for Development of Advanced Diagnostics \protect\\ Hokkaido University Hospital, Hokkaido, Japan}
\affil[f]{Department of Surgical Pathology \protect\\ Hokkaido University Hospital, Hokkaido, Japan}

\authorinfo{*corresponding author}

\begin{document} 
\maketitle

\begin{abstract}
We propose a method to accurately obtain the ratio of tumor cells over an entire histological slide. 
We use deep fully convolutional neural network models trained to detect and classify cells on images of H\&E-stained tissue sections. Pathologists' labels consisting of exhaustive nuclei locations and tumor regions were used to trained the model in a supervised fashion. We show that combining two models, each working at a different magnification allows the system to capture both cell-level details and surrounding context to enable successful detection and classification of cells as either tumor-cell or normal-cell.
Indeed, by conditioning the classification of a single cell on a multi-scale context information, our models mimic the process used by pathologists who assess cell neoplasticity and tumor extent at different microscope magnifications. The ratio of tumor cells can then be readily obtained by counting the number of cells in each class. To analyze an entire slide, we split it into multiple tiles that can be processed in parallel. The overall tumor cell ratio can then be aggregated. We perform experiments on a dataset of 100 slides with lung tumor specimens from both resection and tissue micro-array (TMA). We train fully-convolutional models using heavy data augmentation and batch normalization. On an unseen test set, we obtain an average mean absolute error on predicting the tumor cell ratio of less than 6\%, which is significantly better than the human average of 20\% and is key in properly selecting tissue samples for recent genetic panel tests geared at prescribing targeted cancer drugs. We perform ablation studies to show the importance of training two models at different magnifications and to justify the choice of some parameters, such as the size of the receptive field.
\end{abstract}

\section{INTRODUCTION and PURPOSE}
\label{sec:intro}  

Targeted treatment therapies for various types of cancer rely on DNA analysis of patients' cancer cells to identify the drugs that would benefit them the most. The mutations and fusions of genes that cause cancer are detected with genomics panel tests run on next generation sequencing devices. Input material for these tests comes from several types of biopsies, surgical resections and cell-blocks. The tissue samples are fixed in formalin and embedded in paraffin blocks (FFPE). Genomics panel tests based on gene sequencing operations require a minimum ratio of tumor cells to be present in the analyzed tissue sections to provide accurate results. Typically, a set of slide-mounted FFPE, unstained, about 5-micron thick sections are required for the test. For example, in FDA-approved FoundatioOne \cite{FoundationOne2020} and OncomineDx Target Test \cite{Oncomine2018}, the overall tumor content (ratio) should be more than 20\%. For tissues where the overall tumor content is between 10\% and 20\%, micro-dissection and enrichment should be performed to bring up the overall ratio.  Currently pathologists manually estimate the overall tumor cell ratio from hematoxylin-eosin (H\&E) stained tissue sections, identify areas of high tumor cell ratios and, if needed, micro-dissect tissue fragments to enrich the tumor content to over 20\% \cite{cree2014guidance}. This process is highly manual and subjective and could benefit from automation. Furthermore, pathologists' estimation of the tumor cell ratio has been shown to be inaccurate \cite{smits2014estimation}. In particular, that study showed that 38\% of the estimations would have led to insufficient ratios (less than 20\%), possibly causing false negative results on the genomics panel tests. In addition, these genomics tests are very costly and time-consuming, generally taking a few weeks to complete, and are destructive. Therefore, a method for accurate ratio counting over the entire tumor area is needed. Manually counting individual cells by pathologists does involve counting tens of thousands of individual cells, making it highly impractical. 

\begin{figure} [ht]
\begin{center}
\begin{tabular}{c} 
\includegraphics[width=6.5in]{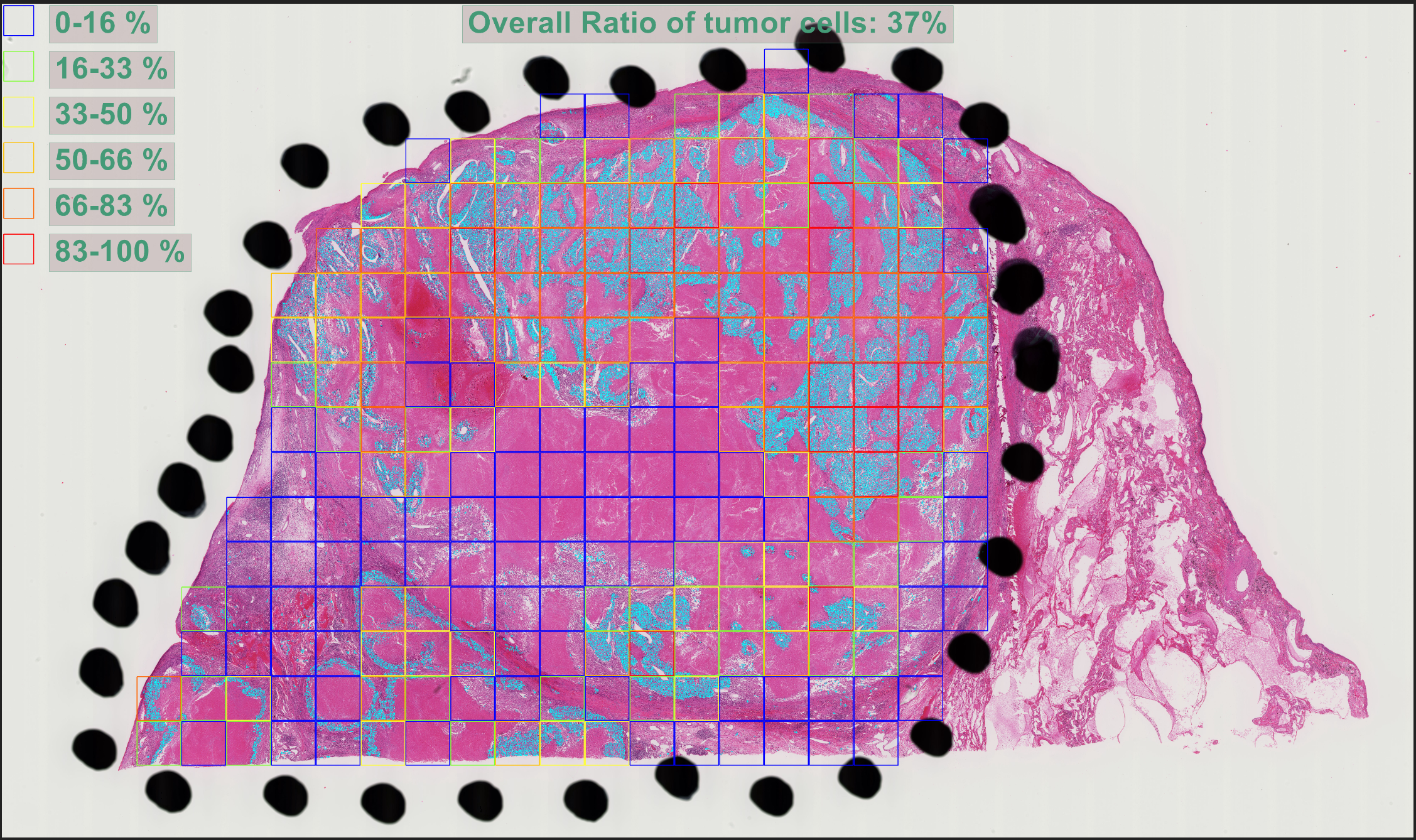}
\end{tabular}
\end{center}
\caption[example] 
{ \label{fig:viewer} 
Screenshot of our browser-based viewer showing the result of analyzing an entire slide of the test set. The overall ratio of tumor cells is shown on top and individual regions are color-coded to visualize the ratios over areas of the slide. Tumor cells are also overlaid in cyan. The black dots are sometimes added by pathologists directly on the glass slide to indicate the tumor area. Our system is able to automatically detect them and analyze the corresponding area. }
\end{figure}

Our goal is to provide an easy-to-use whole-slide interactive system that helps a pathologist select portions of tissue where the ratio of tumor cells is above a safe threshold for use in genomics panel tests. We developed a client-server approach where a browser-based client displays the slide and allows free pan/zoom navigation within the slide. The user submits analysis request for the entire slide or of a marked portion thereof. The AI analysis server processes the requested area and returns the location of all cells and the classification as tumor or non-tumor. The client browser then displays the tumor cell ratio information to the user such that she can decide which tissue area(s) to select for use in genomics panel tests. An example screen shot of the browser-based client tool is shown in Fig.~\ref{fig:viewer}.

\section{PREVIOUS WORK}
\label{sec:previouswork}  

Automated cell counting techniques have been developed in cytology, immuno-histochemistry (IHC) and immuno-fluorescence (IF). In cytology, the most common test is the pap-smear which is automated and was approved by the United States Food and Drug Administration (FDA) in 1998 \cite{bergeron2000quality,tench2002validation}. Cytology images exhibit cells floating in liquid, making the detection and analysis by computer vision systems easier than in tissue sections, where the structure of the tissue interferes with the detection and analysis of the cells. In histology, advanced staining techniques such as IHC and IF can make certain mutated cells easy to segment from the background by using simple color-based analysis techniques. However, such staining is not appropriate for the task of obtaining the ratio of tumor cells because only certain types of mutated cells are highlighted by the stain, while the ratio calculation requires all cells to be counted.

While no cell counting systems exist for H\&E stains that have been FDA approved, several published research studies have addressed the issues of cell detection, classification and segmentation. Two general approaches have been used to detect cells: the first approach starts by detecting candidate objects on the image using image analysis techniques and then use machine-learning techniques, such as support vector machines (SVM) \cite{cortes1995support,cosatto2008grading,arteta2012learning} or K-nearest-neighbors (KNN) \cite{sertel2009computer}, to classify these objects as cells or non-cells using features . Recently, "deep-learning" approaches have taken over and shown to be superior to image-analysis by learning features of cancer from the data instead of relying on "hand-crafted" heuristic features. Such models have been demonstrated to estimate where cells are located on an image, directly, without the need for explicit object detection \cite{lempitsky2010learning, xie2018microscopy, sirinukunwattana2016locality}. Pushing further, some methods directly regress the number of cells present in an image patch \cite{xue2016cell}. Although direct regression has the advantage of bypassing the explicit detection of cells, it is completely black-box and does not provide any way to explain the prediction to the user. We feel that, at a minimum, the user should be able to see which cells are detected and which are classified as tumor cells, so as to gain confidence in the system's predictions. We follow the general object detection approach proposed in Lempitsky et al.\cite{lempitsky2010learning}, teaching a deep convolutional neural network model to learn a mapping between an input image and a density map. The density map is generated from the ground-truth-labeled center of cells' nuclei. 

Most recent cell classification methods employ a deep convolutional neural network (CNN) \cite{lecun1998gradient} trained in a supervised fashion with backpropagation \cite{rumelhart1986learning} to classify a small input image patch into distinct types of cells such as epithelial, fibroblast, mitotic figures \cite{malon2013classification} , etc. Variants of CNN architectures \cite{basha2018rccnet, sirinukunwattana2016locality} have been proposed for this kind of approach. Most methods use image data at a resolution of 0.55 microns per pixel (equivalent to a 20X optical magnification) and perform the analysis on a relatively small receptive field. For example Basha et al.\cite{basha2018rccnet} and  Sirinukunwattana et al. \cite{sirinukunwattana2016locality} use a 32x32 and 27x27 receptive field respectively, which corresponds to about 15 microns, three times the size of a nucleus. Such a small receptive field makes the learning model focus solely on one cell, ignoring the surrounding context. 

Segmentation of tumor areas is another related active research topic. Deep learning methods have been shown to produce excellent results on general image segmentation tasks \cite{he2017mask} and have recently been applied to histology images for nuclei segmentation \cite{naylor2018segmentation, graham2019hover, raza2019micro} and tumor area segmentation \cite{lahiani2019generalising, xu2016deep}. The general principle is the same as for nuclei detection. A model is trained to map an input image to a density map representing the tumor area. For tumor segmentation, the working magnification should be lower than for nuclei detection. While separating individual nuclei requires a high magnification (20X or 40X) and a smaller field of view (50 to 100 microns), segmenting a tumor area necessitates a wider field of view (500 to 1000 microns) with a lower resolution (5X or 10X). Similarly, pathologists observe specimens at low magnification to understand the extent of the tumor area and then zoom in to certain areas to observe finer characteristics such as the morphology of nuclei. We follow this principle in our approach for tumor cell ratio counting. We use a high magnification for detection of nuclei to ensure that all nuclei can be counted, even in dense areas. This is followed by an analysis at low magnification for tumor area segmentation to ensure that large features of the tumor, such as deformed glands, can be seen by the model. In addition, we combine high and low resolution features to classify individual cells as tumor or normal. This allows to properly count normal cells appearing in a tumor area, as well as isolated tumor cells in a normal area. 

Methods using a larger receptive field and a fully-convolutional approach using the U-net \cite{ronneberger2015u} CNN architecture have been demonstrated for cell segmentation in DAPI and FISH staining \cite{rajkumar2019ecseg}. Such methods are ideal of image segmentation as they produce a binary segmentation map directly from the image input. We adapt this segmentation method to cell counting by generating target cell location maps where cells are represented by Gaussian peaks rather than their actual shape. From the predicted map, the (x,y) location of cells is obtained using only local peak detection. This approach avoids the issue of adjacent cells forming clumps that cannot be easily separated into individual cells for counting. We further adapt the method by multi-tasking detection and classification on the same model by learning to predict two target maps, one for detection, where all cells are drawn and one for classification, where only the tumor cells are drawn.

\section{MATERIAL and METHODS}
\label{sec:methods}

\subsection{Data}
\label{sec:methods_data}
One hundred WSI slides were obtained from a cohort of 100 lung cancer cases at Hokkaido University Hospital (Hokkaido, Japan).  Fifty-five cases were categorized as adenocarcinoma (AC) and 45 as squamous-cell carcinoma (SC). The specimens were acquired between 2005 and 2010. Thirty slides contained a tissue micro-array (TMA) specimen (15 AC and 15 SC), while seventy slides contained tissue from surgical resections (40 AC and 30 SC). The tissue specimens were prepared with the Formalin-Fixed Paraffin-Embedded (FFPE) method and stained with Hematoxylin-Eosin (H\&E). The slides were scanned using a Hamamatsu scanner \cite{Hamamatsu2020} into whole slide image (WSI) files. 131 regions of interest (ROI) were selected by a pathologist (KCH) for annotations. These ROIs were 1980x1980 pixels at level-0 (40X) magnification, corresponding to a standard microscope high-power field. Two types of annotations were obtained from a pathologist: point location of the center of each cell’s nucleus and freehand traces of the tumor areas. 

\subsection{System}
\label{sec:methods_system}

We combine two deep neural network (DNN) models and train them in supervised fashion, one to detect and classify cells as normal or tumor, the other to segment tumor areas. Following the work of Lempitsky et al. \cite{lempitsky2010learning}, we train the DNN models to learn a mapping between an input RGB image patch and a target map. 

\begin{figure} [ht]
\begin{center}
\begin{tabular}{c} 
\includegraphics[width=6.5in]{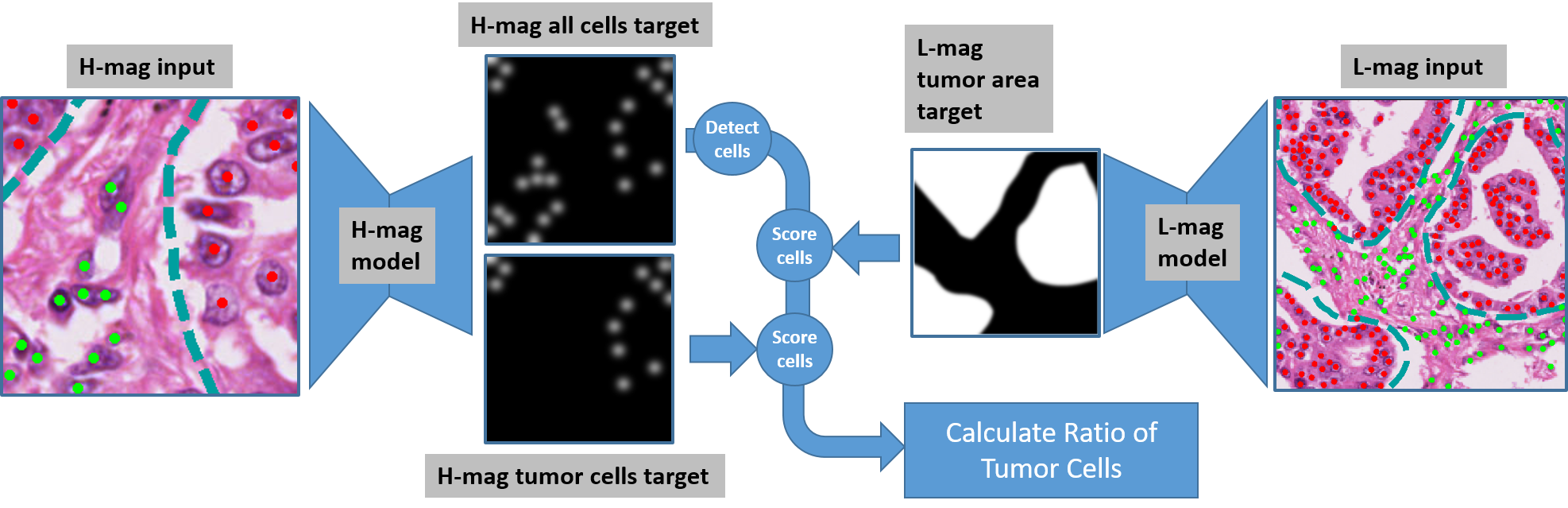}
\end{tabular}
\end{center}
\caption[example] 
{ \label{fig:system} 
System overview. The left part shows the high-magnification process of detecting cells and classifying them as tumor, while the right part shows the low magnification process of segmenting the tumor area. Scores from individual cell classification and tumor area segmentation are combined to obtain a tumor score for each cell. The ratio of tumor cells can then be readily computed. Ground-truth annotations used to generate the targets are shown as overlays on the input images (green dots mark normal cells, red dots mark tumor cells, dotted teal lines mark the tumor areas). }
\end{figure}

For the first model, two target maps are concatenated to form the target output for training the first model. One target map represents the position of the center of \em all \em cells' nuclei present in the input image. This map will be used for nuclei detection. The second target map represents the positions of the \em tumor \em cells' nuclei and is used for nuclei classification. The maps are built by drawing disks at the (x,y) positions of the nuclei center and then by transforming the disks into Gaussian peaks by convolving a Gaussian kernel with the map (see left side of Fig.~\ref{fig:system} for an illustration of the two nuclei maps and their corresponding input image and annotations). The reason for this step is to train the model to produce a smooth peak at a nucleus position, making it easier for post-inference processing to detect them individually, especially in the case where groups of nuclei are bunched-up together. This way, a simple and fast local peak detector checking a 3x3 neighborhood can detect the peaks directly from the DNN output map.

For the second model, the target map is generated in a similar way, at lower magnification, by drawing a filled freehand tumor area against the background. See the right side of Fig.~\ref{fig:system} for an illustration of the tumor area map and its corresponding input image and annotations (dashed contour lines).

\subsection{Model architecture}
\label{sec:methods_model}

Our choice for model architecture is fully-convolutional. These models are the natural choice for detection of multiple objects as they conserve the image's 2D relationships throughout the layers. We experimented with both U-net \cite{ronneberger2015u} and Resnet \cite{szegedy2017inception} with a fully-convolutional head. These models have been applied successfully in a wide variety of image problems. We found that, for our application, U-net has a small performance edge, a simpler architecture and smaller footprint. 

\begin{figure} [ht]
\begin{center}
\begin{tabular}{c|c|c} 
\includegraphics[width=2.6in]{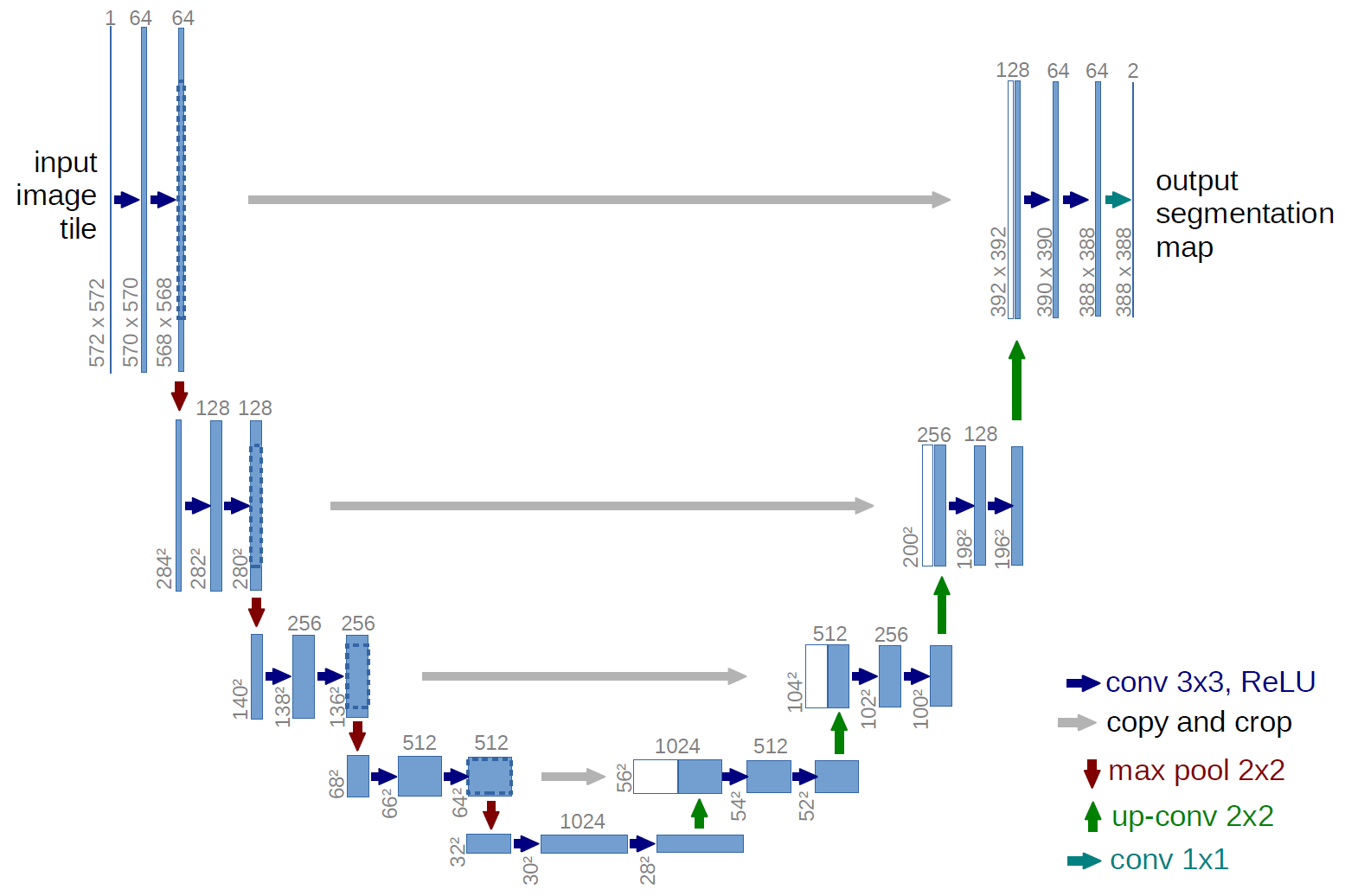} & \includegraphics[width=1.7in]{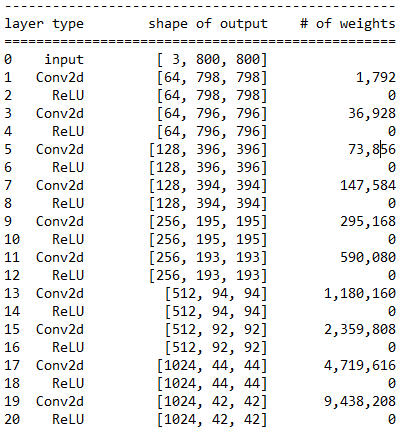} & \includegraphics[width=1.7in]{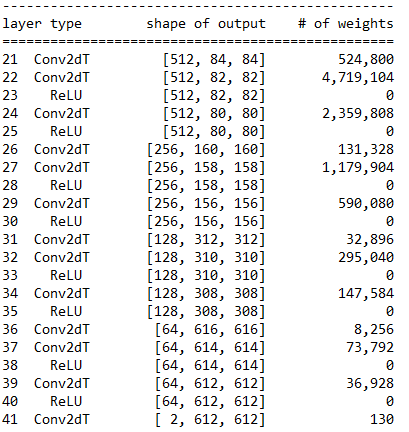}
\end{tabular}
\end{center}
\caption[example] 
{ \label{fig:layers} 
Architecture of the U-net fully-convolutional model. The left image shows the graphical representation of the model (for a 572x572 input image). The middle and right tables show the actual size of the layers for the encoder and decoder block respectively, for an 800x800 input image and 2 output maps. The total number of parameters is 28,942,850 and the total size of the model for inference (not counting training gradients) is about 3GB.  }
\end{figure}

U-net models have an encoder-decoder architecture with a bottleneck in the middle. The number of feature planes increases as their size decreases in the encoder, with the reverse happening in the decoder. The convolutional units in the decoder are implemented with transposed convolutions which can be seen as the gradient of the convolution with respect to its input. A graphical representation of the model is shown in figure \ref{fig:layers}, left.

We choose the number of convolutional blocks such that the model's receptive field is 188x188 pixels, which, at 40X magnification (4.4 pixels per microns), corresponds to a patch of 43x43 microns (172x172 microns at 10X). The receptive field reflects how many input pixels a pixel in the bottleneck 'sees'. We want to make sure that the context seen by the model at 10X magnification is wide enough to include tissue elements such as gland formations and epithelial layers. Conversely, at 40X, we want the model to see individual nuclei with enough details to be able to tell neighboring nuclei apart. Figure \ref{fig:input_aug} shows the receptive fields at 40X and 10X magnification.

We also combine the detection and classification tasks into one model by sharing its a backbone and adding two output maps, one for each task. This multi-task approach is advantageous as the low-level features learned by the model are shared among the two tasks. 

Finally, as the actual input size to the model can be increased at will, this being a \em fully convolutional \em model, we can size it such as to optimize the number of models that can occupy a GPU memory. For example, for a 2 model configuration, an input image of 800x800 pixels allows both models to be loaded in a 1080 GPU (8GB). Individual layers of the U-net model are listed in figure \ref{fig:layers}.

\subsection{Training}
\label{sec:methods_train}

Both the detection/classification model and the segmentation model are trained independently using the same procedure. Their outputs are combined to detect and classify the cells. Instead of jointly optimizing the models, we train them separately and only subsequently optimize the detection and classification thresholds to achieve the lowest possible error on the validation set. This approach is preferable as the input to the two models are different and the loss function may be dominated by one model.

We partition the annotated data into three sets. Seventy percent of the data is used to train the model, ten percent is used for validation of the model and the remaining twenty percent is used only for the final evaluation. These subsets are built such that all annotation ROIs from a given slide go into the \em same \em subset. 

\begin{figure} [ht]
\begin{center}
\begin{tabular}{c|c|c|c} 
\includegraphics[width=1.5in]{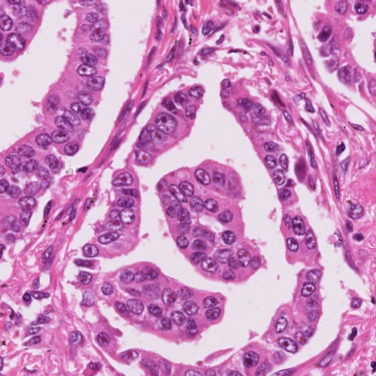} & \includegraphics[width=1.5in]{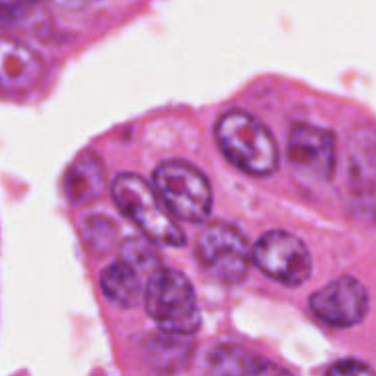} & \includegraphics[width=1.5in]{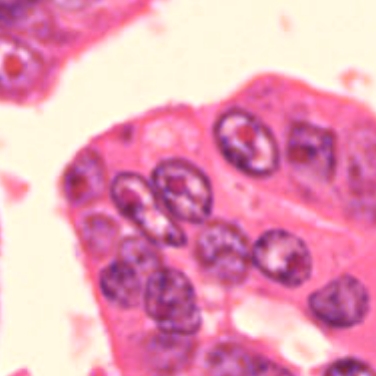} & \includegraphics[width=1.5in]{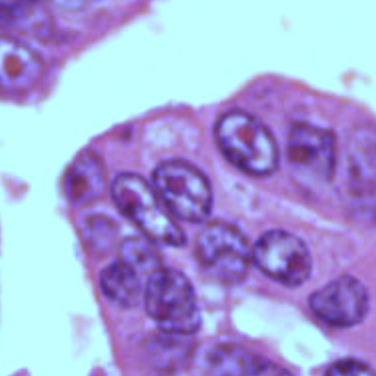} 
\end{tabular}
\end{center}
\caption[example] 
{ \label{fig:input_aug} 
Left side: receptive field of the model at 10X and 40X magnification. Right side: example of data augmentation by H\&E stain shifts. }
\end{figure}

H\&E stained specimens exhibit shades of two staining agents (blue-purple for hematoxylin which colors the nuclei, and reddish-pink for eosin which colors the cytoplasm and extracellular matrix). The amount and proportion of staining agents, the age of the sample and the type of scanner significantly affect the final color of the pixels. Hence it is necessary, in order to create a robust model, to make sure there is as much staining and scanning variation as possible in the training set. Unfortunately, it is difficult to procure samples with such variations, as cohorts tend to come from the same institution and therefore have been stained and scanned using the same protocol. To compensate for this relative uniformity in our training data and to make sure our models will perform adequately on specimens from other institutions, we apply data augmentation to simulate variations in staining, color shifts and  image sharpness. To simulate staining variations, we use an optical-density based stain projection method \cite{ruifrok2001quantification} and shift the pixel intensities in the projection spaces. Figure \ref{fig:input_aug} shows an example of stain augmentation using our method. To simulate color shifts due to the scanner light, we apply intensity shifts in the  Hue-Saturation-Luminance color space. To simulate variations in scanner optics and focusing mechanism, we apply small amounts of pixel blur/sharpen.   Finally, since image orientation does not affect the labels, we also increase the number of examples by random rotation and mirroring.

The model is trained with the binary cross-entropy loss combined with a sigmoid layer $ \sigma(x) $:

\begin{equation}
  \label{eq:twoentropy}
  L_{\mbox{\tiny BCE}} = -\frac{1}{N} \sum_n^N
  \left[ y_n \ln \sigma(x_n) + (1-y_n) \ln (1-\sigma(x_n)) \right]
\end{equation}

where N is the number of pixels in the output map. 

\begin{figure} [ht]
\begin{center}
\begin{tabular}{c} 
\includegraphics[width=4in]{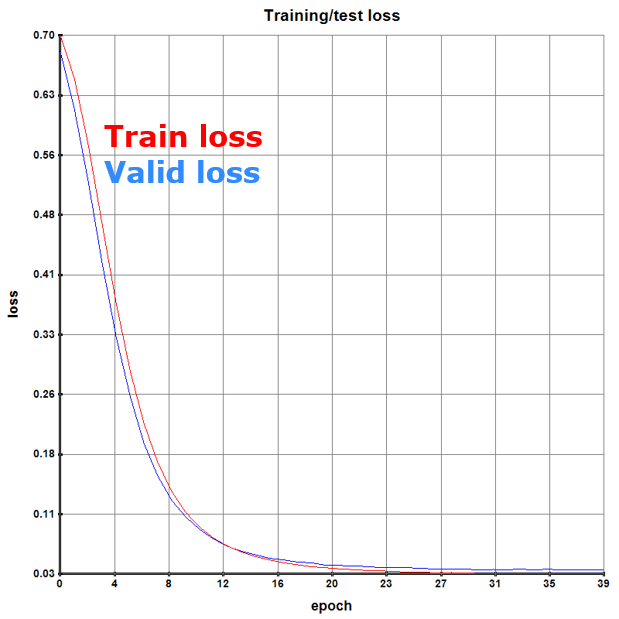}
\end{tabular}
\end{center}
\caption[example] 
{ \label{fig:trn_curve} 
Training curve showing the loss at each epoch. In red is the loss over the training set, while in blue is the loss over the validation set. The training is stopped when the loss on the validation set ceases to decrease. }
\end{figure}

We experimented with batch normalization and found it to be useful in providing a smooth learning curve. We use the Pytorch \cite{paszke2019pytorch} toolchain to train our models using the Adam \cite{zhang2018improved} optimizer and a learning rate of 1e-3. Training a model takes about 3 hours on a GPU. To avoid overfitting the model on the training set, training is stopped when the loss on the validation set stops to decrease for 4 epochs. At each epoch the model is trained with 4000 examples and achieves convergence in about 50 epochs. Figure \ref{fig:trn_curve} shows the training curve of our model.

\subsection{Postprocessing}
\label{sec:methods_eval}

\begin{figure} [ht]
\begin{center}
\begin{tabular}{c|c|c} 
\includegraphics[width=2in]{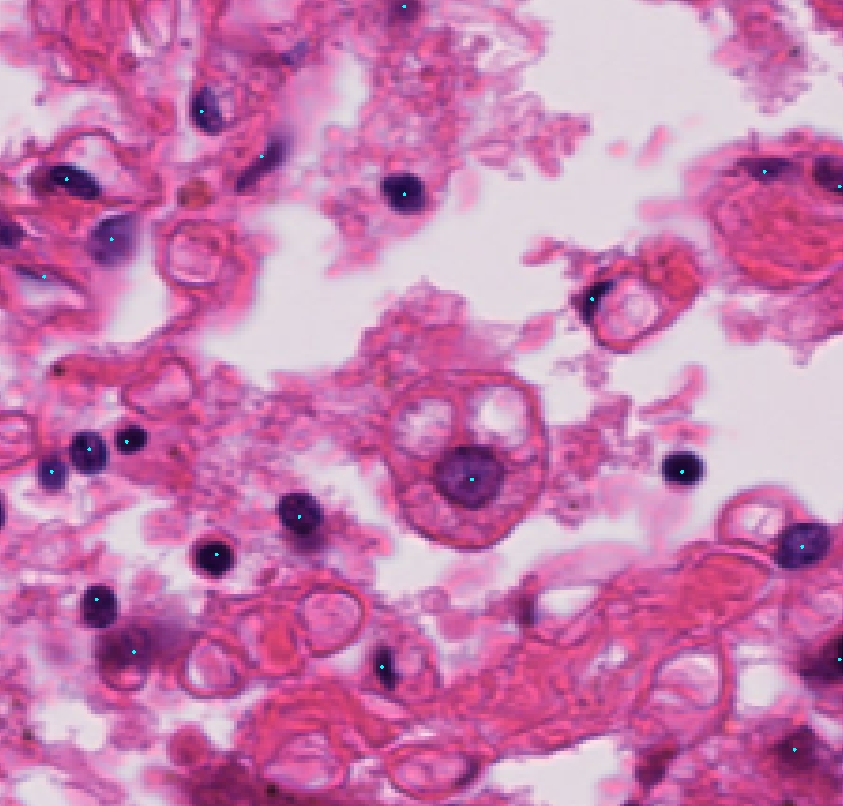} & \includegraphics[width=1.9in]{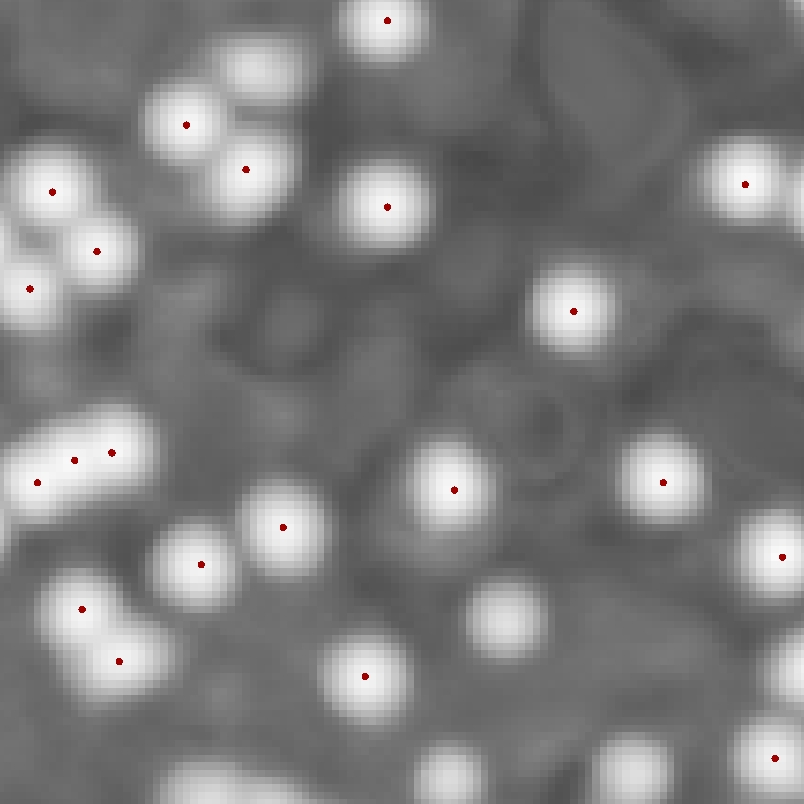} & \includegraphics[width=2.1in]{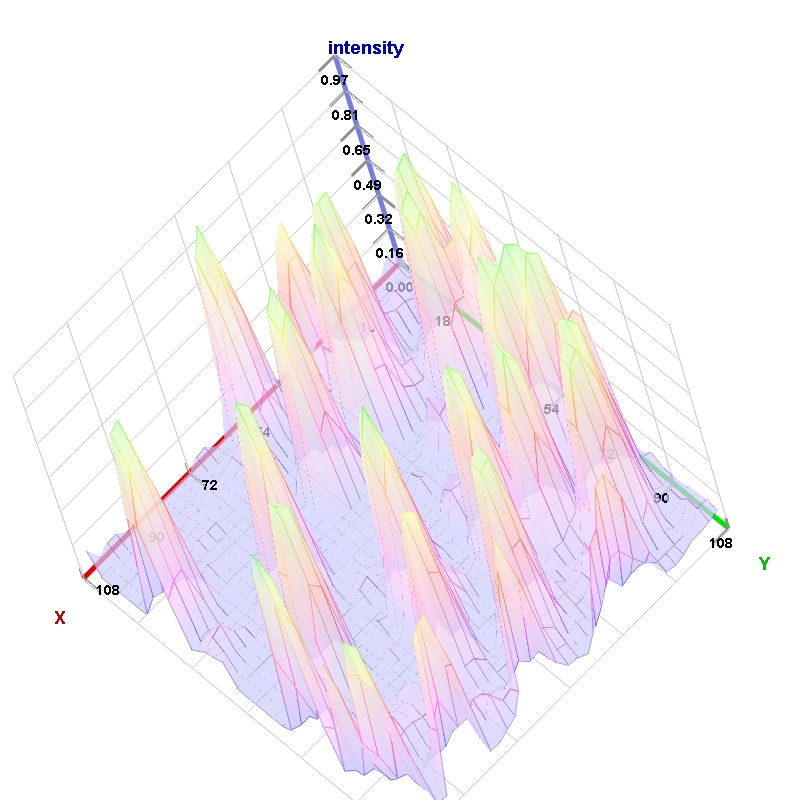}
\end{tabular}
\end{center}
\caption[example] 
{ \label{fig:dmap} 
Detection output map (center) and its 3D view (right) for the input image (left). Locations of peaks are overlaid on the images.  }
\end{figure}

So far the models have been trained individually to minimize the cross-entropy loss, which is a reconstruction loss on the detection, classification and segmentation maps. The real goal, however, is to obtain a list of cells, each with a $(x,y)$ location and a tumor score. The detection/classification model outputs two floating-point maps: $ map_d $ and $ map_c $, while the segmentation model outputs $ map_s $. From $ map_d $, the detection of the cells is performed using local peak detection of 3x3 pixel neighborhood on the detection map (see figure \ref{fig:dmap} center). Since the model is trained to predict smooth peaks, non-maxima suppression is not necessary and every detected peak is considered a candidate cell. The three intensity values at the $ map_k(x,y) $ locations of cell $ i $  is recorded for all three maps resulting in a feature vector $ f_i = [I_d, I_c, I_s] $. 

A classifier is then constructed with two thresholds, the detection threshold $ t_d $ and the classification threshold $ t_c $. Considering the feature vector $ f_i =  [I_d, I_c, I_s] $  of all candidate cells, cells where $ I_d < t_d $ are discarded. Then, cells where $ \alpha  I_c + (1-\alpha)  I_s > t_c $ are classified as tumor cells. $ \alpha $ is a hyper-parameter that weighs the classification and the segmentation model output to generate a score. We use a value of $ 0.5 $.

\subsection{Evaluation}
\label{sec:methods_eval}

The first step is to evaluate the detection accuracy of the detection model. We declare a detected cell a match when the detected peak $(x,y)$ location is within a distance of 3.2 microns (14 pixels at 40X magnification - 4.4 microns per pixel resolution) of a labeled cell. Each detected cell is matched to its closest unmatched label, in a greedy way. Matched cells are true positives (TP), leftover unmatched cells are false positives (FP) and unmatched labels are false negatives (FN). The detection accuracy is thus computed as $ ACC(det) = \frac{TP}{TP+FP+FN} $, the precision as $ PRE(det) = \frac{TP}{TP+FP} $ and the recall as $ REC(det) = \frac{TP}{TP+FN} $. The detection $ F_1 $ score, also known as the Dice coefficient, is defined as $ F_{1}(det) = \frac{2 PRE(det) REC(det)}{PRE(det) + REC(det)} $.

For classification, we calculate the accuracy using the matched detected cells only. A matched cell that is classified as tumor and which label is also tumor is declared a true positive (TP).  A matched cell that is classified as tumor and which label is normal is declared a false positive (FP). A matched cell that is classified as normal and which label is also normal is declared a true negative (TN). A matched cell that is classified as normal and which label is tumor is declared a false negative (FN). The classification accuracy is thus computed as $ ACC(cla) = \frac{TP+TN}{TP+TN+FP+FN} $. For tumor cells (positives), the precision is calculated as $ PRE_{pos}(cla) = \frac{TP}{TP+FP} $ and the recall as $ REC_{pos}(cla) = \frac{TP}{TP+FN} $. For normal cells (negatives), the precision as $ PRE_{neg}(cla) = \frac{TN}{TN+FN} $ and the recall as $ REC_{neg}(cla) = \frac{TN}{TN+FP} $. The classification $ F_1 $ score is defined as $ F_{1}(cla) = \frac{2 P R}{P + R} $ where $ P = \frac{PRE_{pos}(cla)+PRE_{neg}(cla)}{2} $ and $ R = \frac{REC_{pos}(cla)+REC_{neg}(cla)}{2} $ are the average precision and recall over the detected tumor cells and non-tumor cells.

The predicted ratio of tumor cells is defined as $ \widehat{TCR} = \frac{N_T}{N} $ where $ N_T $ is the number cells classified as tumor and $ N $ is total number of detected cells.

\subsection{Threshold tuning}
\label{sec:tune}

We create an evaluation set by combining the training and the validation sets.
Using this evaluation set, we perform a grid search for the best threshold pair $ (t_d,t_c)$. For the detection threshold $t_d$, the criterion we use is to maximize the detection $F_1$ score. For classification, the criterion is to minimize the mean absolute error on the predicted ratio: 

\begin{equation}
  \label{eq:eratio}
  E_{TCR} = \frac{1}{N} \sum_n^N
  \left| \widehat{TCR} - TCR \right|
\end{equation}

Since we use two distinct criterion, each threshold is searched independently: first, the optimal detection threshold, followed by the classification threshold. We found that jointly searching for both thresholds using a single classification criteria yielded worse results and took longer to perform. Once the optimal thresholds are obtained on the evaluation set, the final performance on the test set can be obtained.

\section{RESULTS}
\label{sec:results}

\begin{figure} [ht]
\begin{center}
\begin{tabular}{c} 
\includegraphics[width=6.5in]{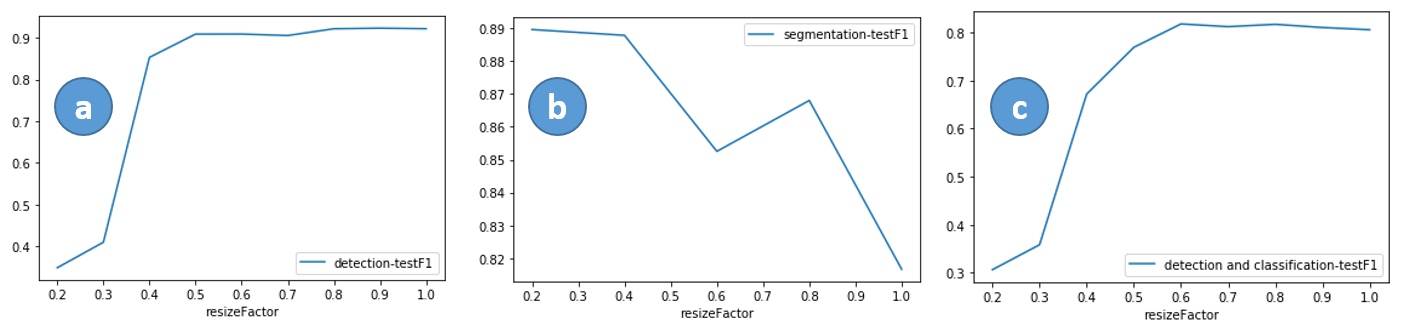}
\end{tabular}
\end{center}
\caption[example] 
{ \label{fig:results} 
Importance of magnification level (x-axis: resizeFactor=1.0 $\equiv$ 40X, resizeFactor=0.2 $\equiv$ 8X) a) Cell detection, b) Tumor Cell classification, c) Cell detection+classification. The y-axes on the plots represent F1 scores (aka Dice Coefficients) on the test set. }
\end{figure}

In a first experiment, we investigate the importance of the magnification level. For the detection of cells, the plot \ref{fig:results}.a) shows that higher magnifications are more accurate, with a dramatic drop in accuracy between magnifications 16X and 12X. This is easy to understand intuitively as the cells become closer and increasingly harder to tell apart at lower magnifications. To show the impact of magnification on the classification of cells, we use ground-truth cell positions and use the tumor cell density output map to classify them as normal or tumor. The trend on plot \ref{fig:results}.b) shows a clear decrease for higher magnifications. The intuition here is that a larger field of view is useful to understand the exact extent of the tumor. We then apply detection and classification, both at the same magnification. The trend on plot \ref{fig:results}.c) shows a maximum at around 24X. This can be understood as a combination of the two previous experiments, exhibiting a sweet spot where each cell's nucleus can still be accurately separated from the others, while providing enough context to classify them as tumor or normal.

In light of the above result and to take full advantage of both the higher detection accuracy at higher magnifications and the higher classification accuracy at lower magnifications, we train two separate models at different magnifications. 
We then combine these models using additive ensembling and report the results in table \ref{tab:results}. The best result of 5.3\% mean absolute error (MAE) on the predicted ratio is obtained by detection at 20X, followed by adding the score from the 20X and 10X tumor cell density maps. This approach can also be seen as conditioning the classification of detected cells on a larger context of tissue. We also show that the U-net architecture returns slightly better results.

\begin{table}[ht]
\caption{Results for 2  model architectures and 2 configurations of the detection/classification (DT+CL) model and the segmentation model (SEG). We report $ E_{TCR} $, the MAE of the predicted tumor cell ratio and the processing speed in $mm^2/sec$. We trained these models on 3 different random partitions of the data and report the mean values and standard deviations.} 
\label{tab:results}
\begin{center}       
\begin{tabular}{|l|c|c|c|} 
\hline
\rowcolor{light-gray}
Resnet50+FC head           & number of model(s) & $ E_{TCR} $                     &   $mm^2/sec$ \\
\hline
DT+CL@20X                    & 1                           & 6.6\% $\pm$1.4                &   1.02             \\
\hline
DT+CL@20X + SEG@10X & 2                           & \textbf{5.8\% $\pm$1.1 }  &  0.47              \\
\hline
\rowcolor{light-gray}
U-net                              & number of model(s) & $ E_{TCR} $                     &   $mm^2/sec$  \\
\hline
DT+CL@20X + SEG@10X & 2                           & \textbf{5.3\% $\pm$1.1 } &   0.46               \\
\hline
DT+CL@10X                    & 1                           & 5.7\% $\pm$1.3              &    3.75               \\
\hline 
\end{tabular}
\end{center}
\end{table}

An important aspect of our system is processing speed. Because the whole slide has to be analyzed to find the most promising areas where the tumor cell ratio is high, a large amount of pixels have to be processed. Our system splits the processing area into rectangular tiles that can be processed independently. We then take advantage of recent multi-core CPU and GPU hardware to accelerate the overall processing. Specifically, given the number of GPU present on the server and the memory footprint of the model, our system optimally spreads processing of tiles to CPU cores and GPUs. Table \ref{tab:results} indicates the time for tested configurations. The fastest configuration uses a single detection/classification U-net model at 10X magnification and produces competitive results. Table \ref{tab:results_f1} shows the performance of this model in terms of accuracy and $F_1$ score and table \ref{tab:timingres} summarizes its processing time on various hardware configurations.

\begin{table}[ht]
\caption{Cell-level accuracy and F1-score for the U-net DT+CL@10X configuration. We report the detection accuracy and F1 score as well as the classification accuracy and F1-score. The classification accuracy and F1-score are obtained on the correctly detected cells. We trained this model on 3 different random partitions of the data and report the mean values and standard deviations. } 
\label{tab:results_f1}
\begin{center}       
\begin{tabular}{|l|c|} 
\hline 
Detection accuracy \% &	92.9 $\pm$ 0.1  \\
\hline 
Detection F1-score \% &	90.4 $\pm$ 0.1  \\
\hline 
Classification accuracy \% &	90.5 $\pm$ 1.7\\
\hline 
Classification F1-score \% &	83.4 $\pm$ 2.4 \\
\hline 
\end{tabular}
\end{center}
\end{table}

\begin{table}[ht]
\caption{Left: speed and acceleration factors for various CPU/GPU hardware setups and for a 1-model DT+CL@10X configuration (see figure \ref{fig:results}). The speed is given in square millimeters of tissue processed per second. A large resection tissue sample may be as large as 600 mm\textsuperscript{2}, while core needle samples are usually about 10 mm\textsuperscript{2}. Right: timing profile for the fastest configuration.} 
\label{tab:timingres}
\begin{center}       
\begin{tabular}{|c|c|c|} 
\hline
\rowcolor{light-gray}
Hardware configuration	& $mm^2/sec$	& speedup \\
\hline 
3 GPUs and 9 CPU cores	& 3.7	& 46 X \\
\hline 
3 GPUs and 6 CPU cores	& 3.3	& 40 X \\
\hline 
2 GPUs and 4 CPU cores	& 2.5	& 30 X \\
\hline 
1 GPU  and 2 CPU cores	& 1.7	& 20 X \\
\hline 
0 GPU and 4 CPU cores	& 0.18	& 2 X \\
\hline 
0 GPU and 1 CPU core	& 0.08	& 1 X \\
\hline 
\end{tabular}
\begin{tabular}{|c|c|} 
\hline
\rowcolor{light-gray}
Operation	& \% of total time \\
\hline 
Read pixels &  42 \\
\hline 
Model inference &  36 \\
\hline 
Save result &  11 \\
\hline 
Peak detect & 7 \\
\hline 
Model setup &  4 \\
\hline 
\end{tabular}
\end{center}
\end{table}

\section{CONCLUSIONS}
\label{sec:conclusions}

We show that conditioning the classification of detected cells on tissue context information provides increased accuracy. Hence, we propose a model that combines high magnification cell detection with low magnification tumor area segmentation, taking advantage of high resolution to accurately separate cells, while using a larger field of view to better classify them as normal or tumor cells. We use fully convolutional DNN architectures and predict density maps from which cell counts can be readily extracted. Finally, we propose a whole-slide approach to calculate the tumor cell ratio by splitting the tissue into square tiles, processing them in parallel and displaying a heatmap of the tumor cell ratio in a browser-based client. The achieved MAE of 5.3\%$\pm$1.1 on the predicted ratio means that a pathologist can safely select areas of the tissue with tumor cell ratios above 27\% (assuming a 20\% cutoff for the genomics panel tests), a clear improvement over human estimation which averages a MAE of 20\% \cite{smits2014estimation}. We are currently collecting an extended dataset from different hospitals to test our system for variations in staining protocols. We expect it to be robust against such variation because of our data augmentation approach during the training of the models, which simulates such variations.

Our main contributions are: 1) a novel multi-scale DNN tumor cell detection and classification model which takes advantage of both high magnification to accurately distinguish individual cells and low magnification to classify them as tumor or normal based on a large tissue context; 2) a novel whole-slide tumor cell ratio counter that is highly accurate at 6\% mean absolute error.

\bibliography{DeepTumorCellRatioCounter} 
\bibliographystyle{spiebib} 

\end{document}